\documentclass[aps,prmaterials,twocolumn,groupeaddress,floatfix]{revtex4-2} 
\usepackage{amsmath}
\usepackage{graphicx} 
\usepackage{dcolumn} 
\usepackage{xcolor}
\usepackage{hyperref}
\usepackage{multirow}
\usepackage{bigstrut}

\hypersetup{colorlinks=true,allcolors=blue}

\newcommand*\ce[1]{\ensuremath{\mathrm{#1}}}	   

\newcommand{\Tc}{$T_{\text{Curie}}$}
\newcommand{\Tchem}{$T_{\text{chem}}$}
\newcommand{\etal}{\emph{et al.\@} }			    

\newcommand{\vect}[1]{\textbf{\textit{#1}}}

\begin{document}
\date{\today}

\title{Magnetochemical effects on phase stability and vacancy formation in fcc Fe-Ni alloys}

\author{Kangming Li, Chu-Chun Fu, Maylise Nastar, Fr\'{e}d\'{e}ric Soisson}
\affiliation{Universit\'{e} Paris-Saclay, CEA, Service de Recherches de M\'{e}tallurgie Physique, F-91191 Gif-sur-Yvette, France}

\author{Mikhail Yu. Lavrentiev}
\affiliation{United Kingdom Atomic Energy Authority, Culham Science Centre, Abingdon, Oxon OX14 3DB, United Kingdom}

\begin{abstract}
We investigate phase stability and vacancy formation in fcc Fe-Ni alloys over a
broad composition-temperature range, via a density functional theory parametrized effective interaction model, which includes explicitly spin and chemical variables. On-lattice Monte Carlo simulations based on this model are used to predict the temperature evolution of the magnetochemical phase. The experimental composition-dependent Curie and chemical order-disorder transition temperatures are successfully predicted. We point out a significant effect of chemical and magnetic orders on the magnetic and chemical transitions, respectively. The resulting phase diagram shows a magnetically driven phase separation around 10-40\% Ni and
570-700 K, between ferromagnetic and paramagnetic solid solutions, in agreement with experimental observations. We compute vacancy formation magnetic free energy as a function of temperature and alloy composition. We identify opposite magnetic and chemical disordering effects on vacancy formation in the alloys with 50\% and 75\% Ni. We find that thermal magnetic effects on vacancy formation are much larger in concentrated Fe-Ni alloys than in fcc Fe and Ni due to a stronger magnetic interaction.
\end{abstract}

\maketitle


\section{Introduction}

Magnetism is an indispensable ingredient for understanding and predicting properties in Fe and Fe-based alloys. It plays a crucial role in phase stability and the bcc-fcc phase transition in Fe~\cite{Soulairol2010,Wrobel2015,Hasegawa1983a,Lavrentiev2010,Ma2017}. In Fe-based alloys, the magnetochemical interplay can lead to a change of the chemical order-disorder transition temperature, local segregation or unmixing tendency~\cite{Rahaman2011,Ruban2008,Schneider2018,Mirebeau2019}. Thermal magnetic effects are also known to have an impact on vacancy properties and atomic diffusion in bcc Fe~\cite{Iijima1988,Huang2010,Wen2016,Schneider2020}.

Effects of magnetism can be obtained via first principles calculations, which are routinely performed in magnetically ordered systems. However, it remains a challenging task to model magnetic excitations and paramagnetism~\cite{Abrikosov2016}. First principles approaches to simulate finite-temperature magnetism in alloys include, for instance, the disordered local moment (DLM) and partial DLM methods~\cite{Okatov2009,Delczeg2012a,Razumovskiy2016,Gambino2018} and the spin-wave method~\cite{Ruban2012}. However, these approaches generally require additional interpolation schemes such as the semi-empirical Ruch model~\cite{Ruch1976} to obtain the temperature evolution of magnetic and energetic properties. Furthermore, they are too computationally expensive for a systematic exploration of the spin-atom configurational space of magnetic alloys. On the other hand, while upperscale atomistic approaches such as spin-lattice dynamics~\cite{Ma2012,Ma2020} and spin-atom Monte Carlo simulations~\cite{Lavrentiev2011,Lavrentiev2014,Lavrentiev2016} provide an efficient way to investigate finite-temperature magnetic effects, it is generally difficult to develop accurate models and potentials for concentrated alloys with the presence of structural defects. 

This work is focused on fcc Fe-Ni alloys, which are the basis of austenitic steels. The alloy with around 50\% and 75\% Ni has a ferromagnetic L1$_0$ and L1$_2$ ordered structure, respectively, at low temperatures. They undergo successive chemical and magnetic transitions with increasing temperature~\cite{Swartzendruber1991,Ohnuma2019}. A strong magnetochemical interplay is expected and can have an impact on phase stability and properties of structural defects.

Phase stability of this system has been extensively investigated experimentally and theoretically~\cite{Swartzendruber1991,Mishin2005,
Cacciamani2010,Bonny2009,Mohri2015,Ohnuma2019,Li2020,Dang1996,Ekholm2010,Vernyhora2010,Lavrentiev2014}. However, thermodynamic measurements of, \textit{e.g.}, activity coefficients and formation enthalpies, were performed only in paramagnetic and chemically disordered alloys~\cite{Swartzendruber1991,Cacciamani2010}. It is difficult to estimate the magnetic contribution to phase stability based directly on experimental information. On the theoretical side, magnetic effects on phase stability of fcc Fe-Ni alloys were studied using model Hamiltonians combined with on-lattice Monte Carlo simulations, showing a significant impact on the chemical order-disorder transition temperatures~\cite{Dang1996,Ekholm2010,Vernyhora2010}. However, the Ising or Heisenberg models adopted in these studies~\cite{Dang1996,Ekholm2010,Vernyhora2010} were developed only for specific compositions, and the composition dependence of magnetic moments as well as the thermal longitudinal spin fluctuations were not taken into account. Recently, a magnetic cluster expansion model was parametrized for the whole composition range of fcc Fe-Ni alloys~\cite{Lavrentiev2014}, but the predicted Curie points of the disordered alloys are found to be much lower than the experimental data.

As the simplest structural defect in metals and alloys, vacancy plays a dominant role in atomic diffusion. Knowledge of vacancy formation properties is thus crucial for the understanding of kinetic processes. From a general point of view, theoretical studies addressing finite-temperature magnetic effects on vacancy formation properties have been focused on metals and extremely dilute Fe alloys~\cite{Huang2010,Ding2014,Sandberg2015,Wen2016,
Gambino2018,Hegde2020,Schneider2020,Li2021}. By contrast, vacancy formation energies in concentrated alloys are often computed with the magnetic ground states~\cite{Guan2020,Zhang2015a,Li2019b,Piochaud2014,Wrobel2017,Manzoor2021,Zhao2016}, or, less commonly, in the ideal paramagnetic state~\cite{Delczeg2012a}. Besides, the investigations of the alloying effects on vacancy formation energies are restricted to either nearly perfect ordered phases~\cite{Mayer1995,Hagen1998,Mishin2000,Woodward2001,DeKoning2002,Rogal2014} or fully random solid solutions~\cite{Guan2020,Zhang2015a,Li2019b,Piochaud2014,Wrobel2017,Manzoor2021,Zhao2016,
Ruban2016,Morgan2020,Zhang2021a}. A continuous and comprehensive modelling of vacancy properties as a function of temperature and hence of chemical and magnetic orders is still missing. It is noted that Girifalco~\cite{Girifalco1964} and Ruch \etal{}~\cite{Ruch1976} proposed to express vacancy formation free energy $G_f$ as a function of order parameter $S$:
\begin{equation}
G_f(S) = (1 + \alpha S^2) \cdot G_f(0)   \label{eq:ruch}
\end{equation}
where $\alpha$ is a system-dependent parameter, $S$ is the chemical or magnetic long-range order parameter, and $G_f(0)$ is the vacancy formation free energy in the chemically or magnetically disordered state. However, these interpolation schemes are not applicable to the alloy systems with simultaneous chemical and magnetic evolutions.

There are few theoretical and experimental studies on vacancy properties in fcc Fe-Ni alloys. Zhao \etal{}~\cite{Zhao2016} used density functional theory (DFT) calculations to obtain the distribution of vacancy formation energies in the ferromagnetic disordered structures with 50\% and 80\% Ni, and compared the DFT results with the predictions from empirical potentials. Caplain and Chambron measured vacancy formation energies in Fe-Ni disordered alloys with 50-94\% Ni using the magnetic anisotropy method~\cite{Chambron1974,Caplain1977}. However, the effects of magnetic and chemical orders on vacancy properties remain largely unexplored experimentally and theoretically. Besides, a comprehensive atomic-scale modelling of the vacancy properties as a function of temperature and hence of chemical and magnetic orders is still missing. 

This study is aimed at elucidating magnetochemical effects on phase stability and vacancy formation in fcc Fe-Ni alloys. We develop a new effective interaction model, which is parametrized on DFT results only and includes explicit chemical and magnetic variables. We treat the magnetic interaction within a generalized Heisenberg formalism~\cite{Lavrentiev2014,Schneider2021a} to account for the dependence of magnetic moments on local chemical composition and the strong longitudinal spin fluctuations in this system~\cite{Li2020,Ruban2007b,Li2021}. The model combined with on-lattice Monte Carlo simulations enables to fully take into account the simultaneous magnetic and chemical evolutions with temperature on the whole composition range of the Fe-Ni alloys.

The paper is organized as follows. Details of DFT calculations, model parametrization and Monte Carlo simulations are given in Sec.~\ref{sec:method}. Phase stability predictions including chemical and magnetic transition temperatures, phase diagram and magnetochemical interplay, are presented in Sec.~\ref{sec:phase_stability}. The temperature and concentration dependences of vacancy formation magnetic free energy are discussed in Sec.~\ref{sec:vac}.

\section{Computational details}
\label{sec:method}

As a first step, we performed DFT calculations presented in Sec.~\ref{sec:dft}. Then, these results were used for the parametrization of the effective interaction model as detailed in Sec.~\ref{sec:EIM}. Finally, several Monte Carlo schemes as described in Sec.~\ref{sec:MC} were employed to study phase stability and vacancy formation properties in fcc Fe-Ni alloys.

\subsection{DFT calculations}
\label{sec:dft}

DFT calculations were performed using the projector augmented wave (PAW) method~\cite{Blochl1994,Kresse1999a} as implemented in the Vienna Ab-initio Simulation Package (VASP) code~\cite{Kresse1993a,Kresse1996e,Kresse1996c}. The generalized gradient approximation (GGA) for the exchange-correlation functional in the Perdew-Burke-Ernzerhof (PBE) parametrization~\cite{Perdew1996a} was employed. $3d$ and $4s$ electrons of Fe and Ni atoms were considered as valence electrons. The plane-wave basis cutoff was set to 400 eV. The Methfessel-Paxton broadening scheme with a smearing width of 0.1 eV was used~\cite{Methfessel1989}. The convergence cut-off for the electronic self-consistency loop was set to $10^{\text{-}6}$ eV. The $k$-point grids were adjusted according to the cell size, to achieve a sampling density equivalent to a cubic unit cell with a $16^3$ shifted grid following the Monkhorst-Pack scheme~\cite{Monkhorst1976a}. Atomic magnetic moments were obtained by an integration of spin-up and spin-down charge densities within the PAW spheres, with a radius of 1.302 \AA{} for Fe and 1.286 \AA{} for Ni.

Random solid solutions were represented by special quasirandom structures (SQSs)~\cite{Zunger1990} with minimized atomic short-range order parameters~\cite{Cowley1950,Martinez2011}. Supercells of various sizes (up to 128 atoms) were used for vacancy-free systems. For vacancy-containing alloys, 108-site and 128-site supercells were used. 

In our previous work~\cite{Li2020}, DFT calculations had been performed in the Fe-Ni alloys with the respective magnetic ground states, in which the atomic positions, the cell shapes and volumes were optimized. In this study, we explored various magnetic states of the fcc Fe-Ni structures with local or global magnetic disorders. For these configurations, the atomic positions were fixed to those in the magnetic ground states while the cell shape and volume were optimized.

\subsection{Effective interaction model}
\label{sec:EIM}

In our previous work, effective interaction models (EIMs) were parametrized for pure fcc Fe and Ni systems respectively~\cite{Li2021a}. In this work, they are unified and extended as a single EIM for the whole composition range of fcc Fe-Ni alloys. The Hamiltonian form is similar to the previous ones used to investigate magnetic properties, phase stability~\cite{Lavrentiev2014,Lavrentiev2020,Tran2020a} and vacancy formation and diffusion properties~\cite{Schneider2020,Schneider2021a} of the Fe-based systems. The current EIM has the following form:
\begin{equation}
 \begin{split}
H = & \sum_{i} \underbrace{ \sigma_{i} \cdot (A_{i}M_{i}^2+B_{i}M_{i}^4+\sum_{j} \sigma_{j} \cdot J_{ij}\vect{M}_i \vect{M}_j)}_\text{magnetic} + \\
& \sum_{i} \underbrace{\sigma_{i} \cdot (\epsilon_{i}+\sum_{j} \sigma_{j} \cdot (V_{ij}+\alpha_{ij}T))}_\text{chemical (nonmagnetic)}
 \label{Eq:e_i}
 \end{split}
\end{equation}
where $i$ denotes the $i$-th fcc lattice site, $\sigma_{i}$ is the occupation variable and is equal to 1 (or 0) for an occupied (or vacant) lattice site, and $\sum_{j}$ is a sum over all the neighbouring sites up to the fourth-neighbour shell. 

In the magnetic part of the Hamiltonian, $M_{i}$ is the local magnetic moment, $A_{i}$ and $B_{i}$ are the on-site magnetic parameters, and $J_{ij}$ are the exchange interaction parameters. 

In the nonmagnetic part, $\epsilon_{i}$ is the on-site nonmagnetic parameter, $V_{ij}$ and $\alpha_{ij}$ are the nonmagnetic interaction parameters, and $T$ is the absolute temperature. Indeed, the impact of vibrational entropies of mixing on the phase stability of fcc Fe-Ni structures is shown to be significant~\cite{Tian2019,Li2020}. In the present rigid-lattice EIM, we choose a rather simple way to incorporate these effects: we introduce the nonmagnetic parameters $\alpha_{ij}$ to account for the vibrational entropies of mixing of the ferromagnetic structures. This simple treatment neglects the possible magnon-phonon coupling, and amounts to integrating the contribution from the fast vibrational degrees of freedom into the nonmagnetic pair interactions. The nonmagnetic interactions thus become the pair free energies~\cite{Levesque2011,Senninger2014a,Wang2020,Trochet2021}, instead of the simple pair energies of the usual models, due to the inclusion of the entropic contribution. We are aware that the characteristic time scales of the magnon and phonon excitations may not be very different~\cite{Abrikosov2016,Dutta2020}. In addition, our model does not correctly capture vibrational entropy of vacancy formation. A more sophisticated treatment for vibrational degree of freedom, and for both vacancy-free and vacancy containing systems is beyond the scope of the present study.

The effects of the presence of a vacancy are incorporated as the dependence of the model parameters on the distance from the vacancy~\cite{suppl_mat}. The parametrization procedure and the resulting parameters can be found in the Supplemental Material~\cite{suppl_mat}.

\subsection{Monte Carlo simulations}
\label{sec:MC}

Temperature-dependent properties are determined from the EIM combined with on-lattice Monte Carlo (MC) simulations, using 16$^3$ fcc unit cells containing 16384 lattice sites. Some of these properties are defined as follows. 

All the alloy concentrations are expressed in the Ni atomic fraction. Following the Warren-Cowley formulation~\cite{Cowley1950}, the atomic short-range order (ASRO) parameter for the $n$-th coordination shell is calculated as follows:
\begin{align}
\text{ASRO}_n &= 1 - \frac{x^n_\text{Ni}}{x_\text{Ni}}
\end{align}
where $x_\text{Ni}$ is the nominal Ni concentration, and $x^n_\text{Ni}$ is the average local Ni concentration in the $n$-th coordination shell of Fe atoms. The atomic long-range order (ALRO) parameter for L1$_0$-FeNi and L1$_2$-FeNi$_3$ is defined as
\begin{equation}
\text{ALRO} = \frac{N^{\text{Fe}}_\text{Fe}}{N_\text{Fe}} -  \frac{N^{\text{Ni}}_\text{Fe}}{N_\text{Ni}}
\end{equation}
where $N_\text{Fe}$ and $N_\text{Ni}$ are the total numbers of Fe and Ni, respectively, and $N^{\text{Fe}}_\text{Fe}$ and $N^{\text{Ni}}_\text{Fe}$ are the numbers of Fe in the Fe and Ni sublattices, respectively. The Curie temperature \Tc{} is estimated as the inflection point of the following function~\cite{Lavrentiev2020} fitted to the obtained magnetization values:
\begin{equation}
\label{eq:fit_mag}
\frac{M(T)}{M(T=1 K)} = (1-aT)\frac{1+\text{exp}(-\frac{b}{c})}{1+\text{exp}(\frac{T-b}{c})}
\end{equation}

For vacancy-free systems, we use three types of MC schemes for different purposes: spin Monte Carlo (SMC), spin-atom canonical Monte Carlo (CMC) and semi-grand canonical Monte Carlo (SGCMC)~\cite{Trochet2021}. In SMC simulations, the atomic configuration is fixed while the magnetic configuration evolves with temperature. SMC simulations are used to obtain magnetic properties (\textit{e.g.} magnetization, MSRO and \Tc{}) for a fixed atomic configuration. In CMC simulations, the chemical composition is fixed, while the atomic and magnetic configurations are equilibrated. CMC simulations allow to determine magnetic properties, the ALRO and ASRO parameters of the equilibrium phase for a given Ni concentration and temperature. SGCMC simulations are used as a convenient way to construct the phase diagram and its principle can be found in Ref.~\cite{Kofke1988,Wang2020,Trochet2021}. 

In principle, quantum statistics should be used for the magnetic degree of freedom below the magnetic transition temperature. This has been previously done for the pure systems~\cite{Wen2016,Bergqvist2018,Schneider2020,Li2020}, but a systematic application to alloys can be quite complicated and involve further approximations. In this study, classical statistics is used to control the magnetic and chemical evolutions in alloys for the whole composition range.

For vacancy-containing systems, we compute the vacancy formation free energy $G_f$, which is linked to the equilibrium vacancy concentration $[V]_{eq}$ via:
\begin{equation}
[V]_{eq}=\text{exp}(-\frac{G_f}{k_{\text{B}}T})
\end{equation}
where $G_f$ can include all the non-configurational entropic contributions. To evaluate $G_f$, we use a Widom-type MC scheme~\cite{Widom1963,Sindzingre1987,Li2021thesis}, in which the free-energy difference $G_1-G_0$ between the system 1 and 0 is computed as
\begin{equation}
\label{eq:G_f}
G_1-G_0 = -k_\text{B}T \ln < \exp(- \frac{E_1-E_0}{k_\text{B}T} ) >_0
\end{equation}
where $E_1$ and $E_0$ are the energies of the two systems in the same microstate, and $<...>_0$ denotes the ensemble average of $\exp(- \frac{E_1-E_0}{k_\text{B}T} ) $ in the system 0. If there is the same number of atoms in the system 1 as in the system 0, and if there is one vacancy in the system 1 and no vacancy in the system 0, the free energy difference from Eq.~\ref{eq:G_f} is equal to $G_f$. The details of the scheme is given in the Supplemental Material~\cite{suppl_mat}. 

We recall that our EIM includes only the vibrational entropy of mixing for the phase stability prediction, but it does not allow to predict the vacancy formation vibrational entropy. Consequently, the $G_f$ predicted by our EIM is actually the vacancy formation magnetic free energy $G^\text{mag}_f$ which includes the vacancy formation entropy associated with magnetic and chemical excitations but does not include that of the lattice vibration.

\section{Phase stability of fcc Fe-Ni alloys}
\label{sec:phase_stability}

This section is focused on the phase stability prediction from the EIM for vacancy-free fcc Fe-Ni alloys. First, we validate our EIM by comparing its predictions to experimental data. We also compare the vibrational and magnetic entropic contributions to the chemical transitions. Then, we discuss the calculated fcc phase diagram with other theoretical results. In the last subsection, we elucidate the interplay between magnetic and chemical degrees of freedom.

\subsection{Magnetic and chemical transition temperatures}

In the following, we evaluate the accuracy of the EIM by a comparison of the predicted magnetic and chemical transition temperatures with the experimental values. A comparison of ground-state magnetic, energetic properties and vibrational entropies of mixing between the EIM and DFT results is also given in the Supplemental Materials~\cite{suppl_mat}. 

\begin{figure}
	\includegraphics[width=0.8\linewidth]{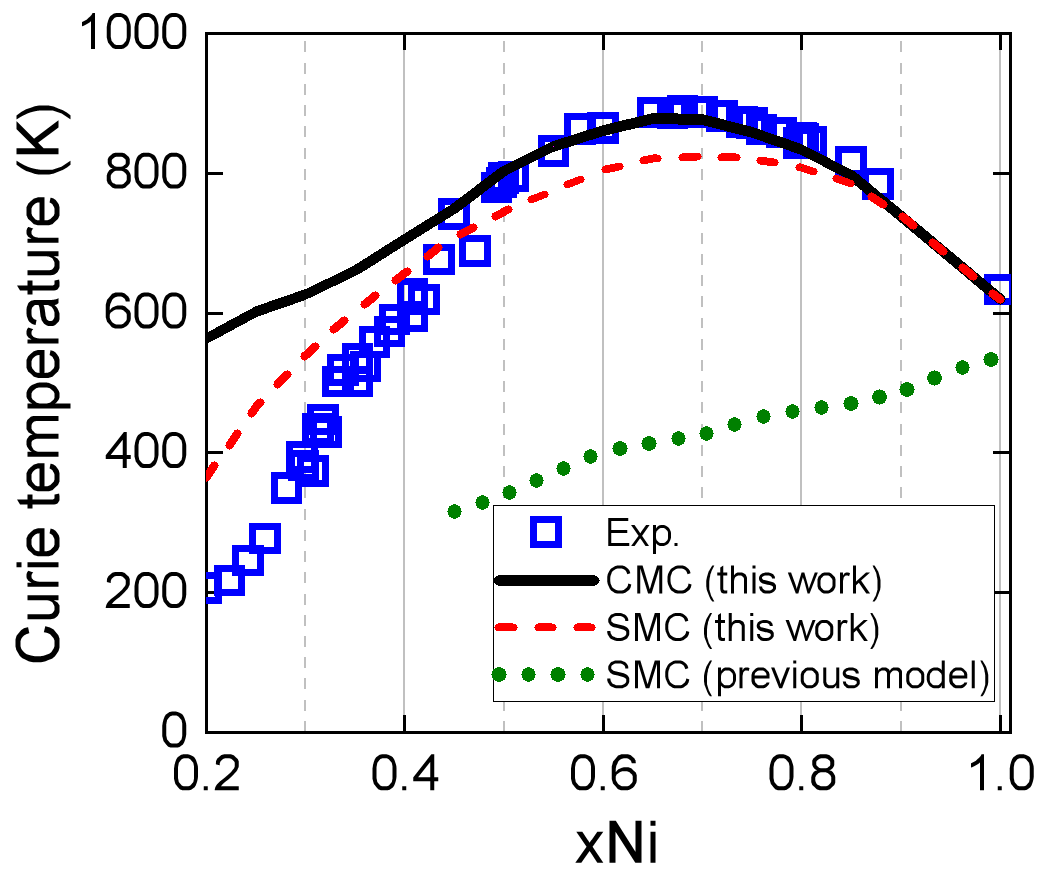}
	\caption{\label{fig:T_curie}(Color online) \Tc{} of fcc random solid solutions from the experiments~\cite{Peschard1925,Crangle1963,Asano1969}, the current EIM and the previous model in Ref.~\cite{Lavrentiev2014}. }
\end{figure}

According to the EIM, the fcc random solid solutions with more than 20\% Ni have a collinear FM ground state~\cite{suppl_mat}. Fig.~\ref{fig:T_curie} shows the predicted and experimental \Tc{}. The experimental \Tc{} were measured in the samples quenched from 923-1273 K~\cite{Crangle1963,Asano1969} with non-zero ASRO. Our SMC results of \Tc{} are obtained with the fully random structures (ASRO equal to zero), whereas the CMC results are obtained with the equilibrium spin-atom structures which have stronger ASRO than the experimental samples. The predicted Curie temperatures of the experimental samples should therefore lie between the CMC and SMC curves.

In alloys with $x_\text{Ni}>0.45$, the CMC results of \Tc{} are in very good agreement with the experimental data in Fig.~\ref{fig:T_curie}, while the SMC results are slightly lower. This indicates that the experimental ASRO is closer to that of the equilibrium structures obtained in CMC simulations than the zero ASRO of the random alloys. However, the CMC results show a large deviation from the experimental data in alloys with $x_\text{Ni}<$0.4. Indeed, the predicted equilibrium structures around 10-40\% Ni at 570-700 K consists of two different disordered phases, as will be shown in Sec.~\ref{sec:phase_diagram}. Therefore, the structures from CMC simulations do not correspond to the experimental homogeneous disordered samples. Meanwhile, the difference between the SMC results and experimental \Tc{} in alloys with $x_\text{Ni}<$0.3 may be due to the non-zero ASRO in the measured samples. Also, our model may describe less well the energetic properties in the alloys very rich in Fe. 
 
The ordered structures L1$_0$-FeNi and L1$_2$-FeNi$_3$ have a FM ground state, with the experimental\Tc{} higher than those in the disordered alloys of the same compositions. This point is well reproduced by the EIM predictions, which compare favourably with the experimental results as shown in Table~\ref{tab:Tcurie_ordered}.

\begin{table}[htbp]
  \centering
  \caption{Comparison of \Tc{} of L1$_0$-FeNi and L1$_2$-FeNi$_3$ between the EIM prediction from SMC simulations and the experiments.}
    \begin{ruledtabular}
    \begin{tabular}{lll}
          & This work & Exp. \\
          \hline
    L1$_0$-FeNi & 845 K & 840 K~\cite{Nagata1986} \\ 
    L1$_2$-FeNi$_3$ & 968 K & 954 K~\cite{Wakelin1953}, 940 K~\cite{Kollie1973}  \\
    \end{tabular}
    \end{ruledtabular}
  \label{tab:Tcurie_ordered}%
\end{table}%

\begin{figure}
	\includegraphics[width=\linewidth]{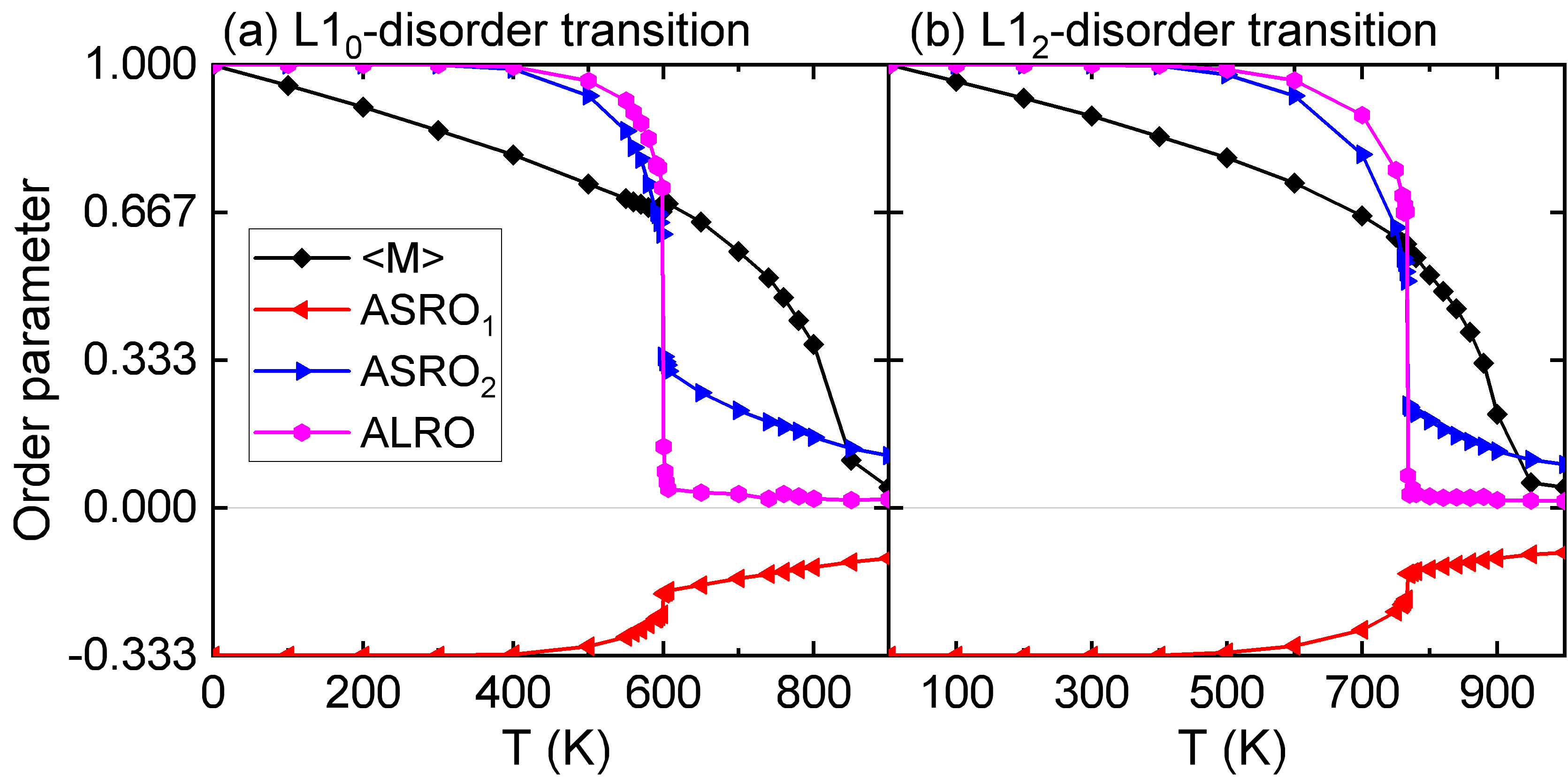}
	\caption{\label{fig:order-disorder}(Color online) Predicted temperature evolution of the reduced magnetization, the ASRO (of the first two shells) and ALRO parameters in the alloys with (a) 50\% and (b) 75\% Ni.}
\end{figure}

The chemical order-disorder transition temperatures \Tchem{} at 50\% and 75\% Ni are obtained from the CMC simulations. As shown in Fig.~\ref{fig:order-disorder}, the ALRO parameter changes abruptly around 598 K and 766 K at 50\% and 75\% Ni, respectively, in excellent agreement with the experimental \Tchem{} of 593 K at 50\% Ni~\cite{Pauleve1962} and of 770-790 K at 75\% Ni~\cite{Kollie1973,Drijver1975,Deen1981}. Above \Tchem{}, the equilibrium structures are found to still retain a degree of ASRO. 
	
We find that \Tchem{} at 50\% and 75\% Ni are increased by 332 K and 154 K, respectively, if the vibrational contribution is switched off in the EIM. This confirms the strong vibrational effects on the chemical transitions in fcc Fe-Ni alloys as suggested in our previous DFT study~\cite{Li2020}. As shown in Table~\ref{tab:order-disorder}, the previous DFT study showed that considering only the ideal configurational entropy leads to a largely overestimated values of \Tchem{}, whereas a reasonable estimation of \Tchem{} can be obtained if vibrational entropies of mixing are included. The effects of magnetic excitations, which are neglected in the previous DFT study~\cite{Li2020} but are accounted for in the EIM, are found to have a smaller impact than the vibrational contribution.


\begin{table}[htbp]
  \centering
  \caption{\label{tab:order-disorder}Chemical order-disorder transition temperatures (in K) in the alloys with 50\% and 75\% Ni. The contributions considered in the calculations are indicated in the parentheses.}
      \begin{ruledtabular}
    \begin{tabular}{lrr}
          & 50\% Ni & 75\% Ni \\
          \hline
    DFT~\cite{Li2020} (conf) & 920   & 1030 \\
    DFT~\cite{Li2020} (conf+vib) & 640   & 830 \\
    EIM, this work (conf+mag) & 930   & 920 \\
    EIM, this work (conf+vib+mag) & 598    & 766 \\
    Exp.~\cite{Pauleve1962,Kollie1973,Drijver1975,Deen1981}  & 593  &  770-790 \\
    \end{tabular}%
        \end{ruledtabular}
\end{table}%

One of the motivations of developing the present EIM is to improve the phase stability prediction of the previous fcc Fe-Ni model in Ref.~\cite{Lavrentiev2014}. For instance, the predicted \Tc{} from this previous model in the disordered structures show a linear dependence on alloy composition and are lower than the experimental data, as shown in Fig.~\ref{fig:T_curie}. In addition, \Tchem{} in the structures with 50\% and 75\% Ni concluded in Ref.~\cite{Lavrentiev2014} are higher than the corresponding \Tc{}, in contradiction with the experimental results. These inconsistencies are fixed in the present EIM, which gives a correct prediction of both the magnetic and chemical transition temperatures.

\subsection{Fcc Fe-Ni phase diagram}
\label{sec:phase_diagram}

\begin{figure*}
	\includegraphics[width=0.85\linewidth]{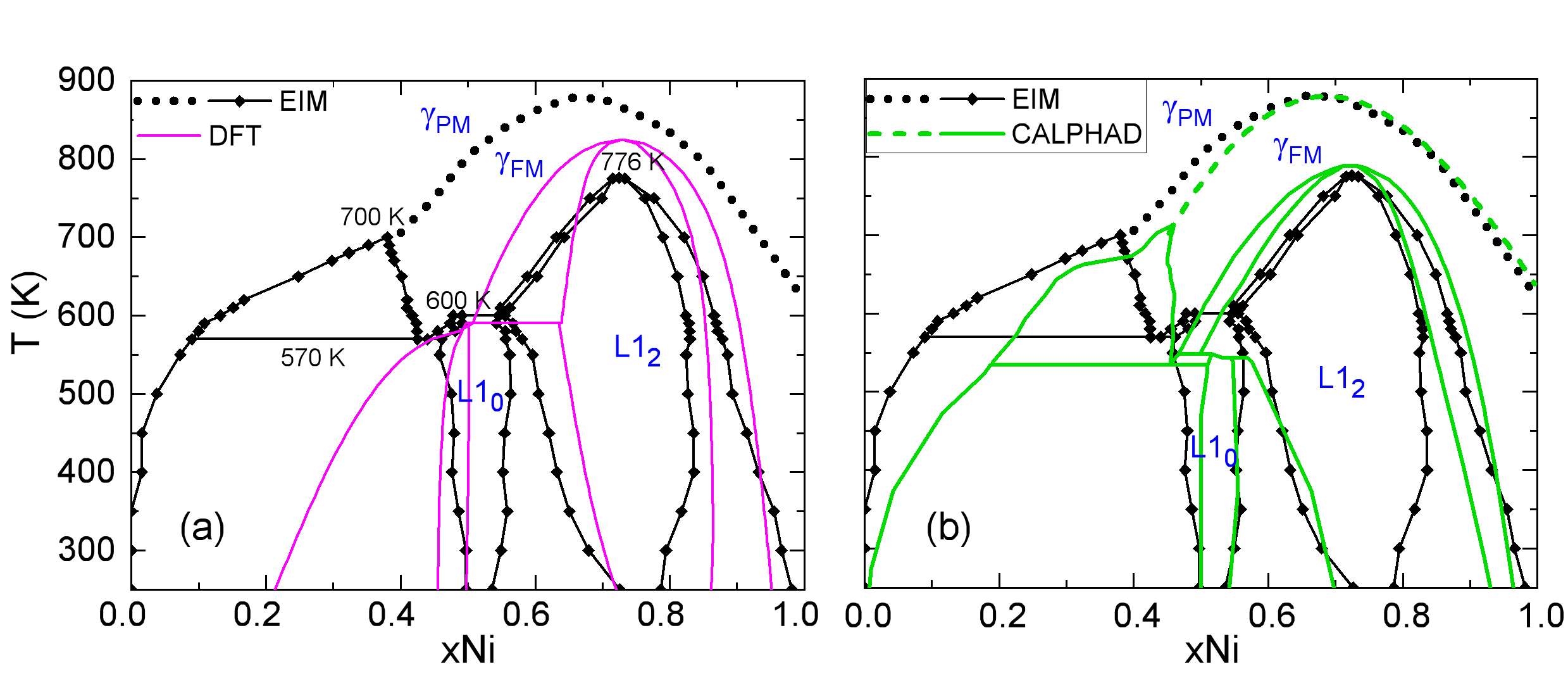}
	\caption{\label{fig:phase_diagram}(Color online) Fcc Fe-Ni phase diagram predicted by the present EIM, compared to the ones from (a) our previous DFT study~\cite{Li2020} and (b) the CALPHAD study by Ohnuma \etal~\cite{Ohnuma2019}.}
\end{figure*}

The phase diagram is constructed by means of the SGCMC simulations. Fig.~\ref{fig:phase_diagram} shows the fcc Fe-Ni phase diagram predicted by the EIM, compared with those from the DFT and CALPHAD studies~\cite{Li2020,Ohnuma2019}. In the following, we denote L1$_0$ and L1$_2$ as the ordered phases around 50\% and 75\% Ni, respectively, and $\gamma_\text{FM}$ and $\gamma_\text{PM}$ as the FM and PM solid solutions, respectively.

According to the EIM prediction, the phase diagram below 570 K consists of four monophasic regions (Fe-rich $\gamma_\text{PM}$, L1$_0$, L1$_2$ and Ni-rich $\gamma_\text{FM}$), which are separated by three corresponding biphasic regions. From 570 to 600 K, the biphasic region $\gamma_\text{PM}$+L1$_0$ is replaced by the biphasic regions $\gamma_\text{FM}$+L1$_0$ and $\gamma_\text{PM}$+$\gamma_\text{FM}$, which disappear at 600 K and 700 K, respectively. The L1$_0$- and L1$_2$-disorder transitions at 50\% and 75\% Ni occur at 600 and 776 K, respectively.

The major difference between the EIM-predicted phase diagram and the DFT one~\cite{Li2020} is the absence of $\gamma_\text{PM}$ in the latter, which considered fully FM phases only. On the other hand, there is no significant difference in the other parts of the two phase diagrams involving the ordered phases. This is not surprising considering the high Curie temperatures of the ordered phases, which remain FM up to the order-disorder transition temperatures.

Recently, Ohnuma \etal{}~\cite{Ohnuma2019} determined experimentally the phase equilibria in Fe-Ni alloys between 673 K and 973 K and revised the thermodynamic descriptions in the CALPHAD modelling. In particular, the \ce{L1_0}-disorder transition temperature is predicted to be 550 K using the revised CALPHAD parameters, in better agreement with the experimental value of 593 K~\cite{Pauleve1962,Reuter1988} than the previous CALPHAD prediction of 313 K by Cacciamani \etal{}\cite{Cacciamani2010}. The fcc phase diagram calculated with the revised CALPHAD parameters of Ohnuma \etal{}~\cite{Ohnuma2019} is presented in Fig.~\ref{fig:phase_diagram}. Despite some differences in the phase boundaries involving the paramagnetic phase, the calculated phase diagrams from EIM and CALPHAD are overall similar. Both predict a small two-phase region between $\gamma_\text{FM}$ and \ce{L1_0}, and a triangle-shape miscibility gap between the ferromagnetic and paramagnetic random alloys. The miscibility gap is consistent with the observations of chemical and magnetic clusters in the Invar alloys~\cite{Chamberod1978,Takeda1985,Takeda1987,Rancourt1987}, in which the Ni-rich and Fe-rich local regions are suggested to be ferromagnetic and paramagnetic respectively~\cite{Takeda1985,Takeda1987}. This miscibility gap will be discussed in more details in the next subsection.

\subsection{Interplay between chemical and magnetic orders}

Magnetization is known to have an impact on the chemical order-disorder transition temperature~\cite{Ekholm2010,Rahaman2011}. To study how different magnetic states influence the chemical transitions, we control the magnetic state with a temperature $T_\text{spin}$ different from the temperature controlling the chemical evolution. To do so, we adopt the adiabatic approximation for the magnetic degree of freedom, namely assuming that the magnon excitations are faster than the chemical evolution. Here we consider two extreme cases for the magnetic state, namely the magnetic ground state and the PM state. 

Table~\ref{tab:diff_Tspin} shows the chemical transition temperatures in the alloys with 50\% and 75\% Ni with different magnetic states. In the alloy with 75\% Ni, the predicted transition temperature ranges from 715 K to 885 K depending on the magnetic state of the system. A strong ferromagnetic order as in the magnetic ground state tends to further stabilize the ordered alloy over the disordered one, while the paramagnetic order reduces the phase stability of \ce{L1_2}-\ce{FeNi_3}. On the other hand, the trend is reversed in the alloy with 50\% Ni. In addition, the influence of the magnetic state on the L1$_0$-disorder transition temperature is less important than on the L1$_2$-disorder transition temperature.

\begin{table}[htbp]
  \centering
  \caption{\label{tab:diff_Tspin}Chemical order-disorder transition temperatures (in K) in the alloys with 75\% and 50\% Ni, obtained with different magnetic states. EQ: equilibrium magnetic state. GS: magnetic ground state within the adiabatic approximation ($T_\text{spin}$=1 K). PM: paramagnetic state within the adiabatic approximation ($T_\text{spin}$=1500 K). }
  \begin{ruledtabular}
    \begin{tabular}{cccc}
    Composition & EQ & GS  & PM    \\
    \hline
    75\% Ni & 766 & 885 & 715  \\
    50\% Ni & 598 & 555 & 610  \\
    \end{tabular}
    \end{ruledtabular}
\end{table}

We have shown that there is a phase separation in the phase diagram around 10-40\% Ni and 570-700 K. To observe the phase separation in a canonical system, the equilibrium structure for a given composition is obtained from CMC simulations. The compositions of the coexisting phases are then estimated from the distribution of local Ni concentration~\cite{Lavrentiev2020}, which is computed for each fcc lattice site as the atomic fraction of Ni atoms within the fifth coordination shell. 

\begin{figure}[htbp]
	\centering
	\includegraphics[width=0.49\linewidth]{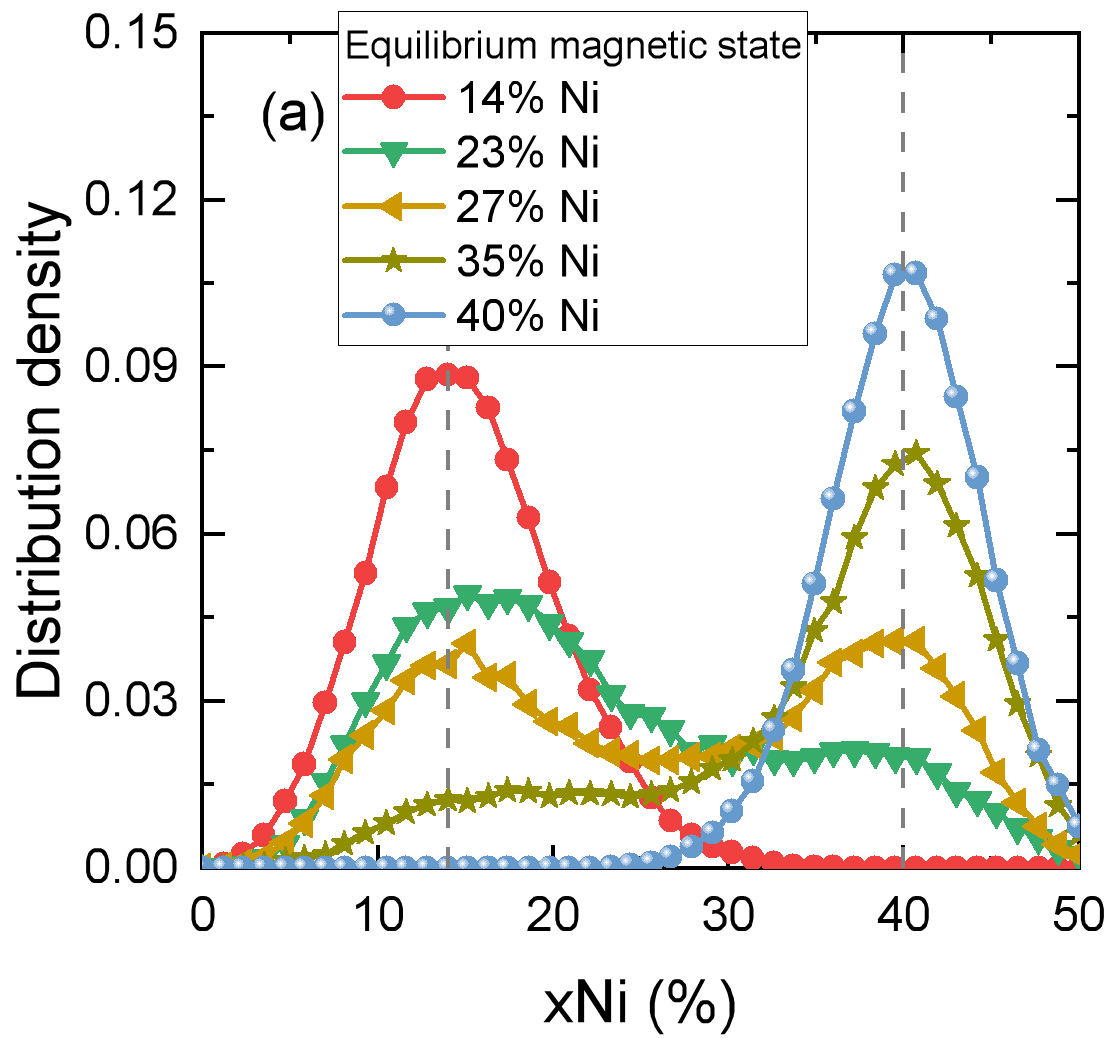}
	\includegraphics[width=0.49\linewidth]{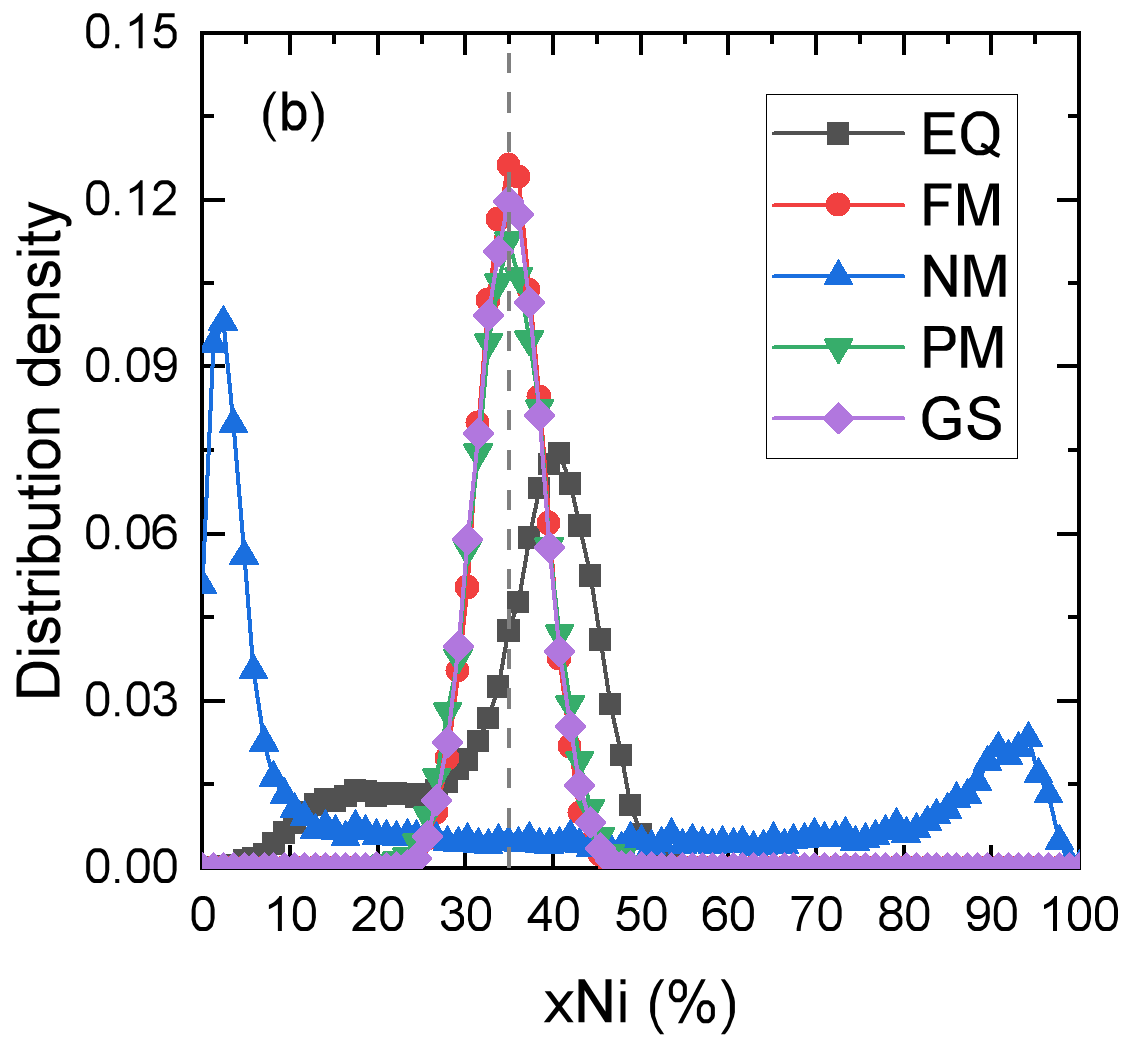}
	\caption{Distribution of local Ni concentration at 600 K. (a) Equilibrium structures with various Ni content, without constraining the magnetic state. (b) Equilibrium structures with 35\% Ni. EQ: equilibrium magnetic state. FM: ferromagnetic. NM: nonmagnetic. PM: paramagnetic within adiabatic approximation ($T_\text{spin}$=1500 K). GS: magnetic ground state within adiabatic approximation ($T_\text{spin}$=1 K).}
	\label{fig:C3__EIM_distr}
\end{figure}

Fig.~\ref{fig:C3__EIM_distr}(a) shows such distributions in the equilibrium spin-atom structures at 600 K. According to our computed phase diagram, the two-phase composition range at 600 K is between 14\% and 40\% Ni, as indicated by the vertical lines in Fig.~\ref{fig:C3__EIM_distr}(a). For the equilibrium structures with 14\% and 40\% Ni, we observe a single peak centred on the nominal composition, which is the signature of a homogeneous single-phase system. At intermediate concentration, the distribution exhibits two peaks indicating the compositions of the two separated phases, namely 14\% and 40\% Ni.

The phase separation may be chemically driven, with the magnetic state simply following the composition of the separated phase, or it may be magnetically driven. To elucidate this point, we study the phase equilibrium in the coexistence region by constraining the magnetic state of the system. Four types of constraints are considered, namely the FM state, the nonmagnetic (NM) state, the magnetic ground state (GS), and the fully paramagnetic (PM) state. Fig.~\ref{fig:C3__EIM_distr}(b) presents the resulting distributions of local Ni concentration in the equilibrium structures with nominal 35\% Ni at 600 K. The distributions obtained in the FM, GS and PM states exhibit one single peak at the nominal concentration, while those obtained in the equilibrium magnetic state and the NM state exhibit two peaks but at different locations. Thereby, there are three different equilibrium atomic states. The equilibrium atomic structure in the NM state is practically a phase separation between fcc Fe and Ni, in line with the positive mixing energies of the nonmagnetic fcc solid solutions. From these results, we conclude that the phase separation between $\gamma_\text{PM}$ and $\gamma_\text{FM}$ is driven by the magnetic interactions.

On the other hand, the magnetic properties of the alloy with a given composition depend on both the atomic long-range and short-range orders (ALRO and ASRO respectively). In order to perform a quantitative analysis of such effects, we extract from the CMC simulations ten different chemical configurations of the 75\% Ni alloy. These structures are representative of the perfect L1$_2$ ordered structure and the fully random alloy, as well as other intermediate states. For these atomic configurations, we run SMS simulations to equilibrate the magnetic state and measure the corresponding Curie temperature.

\begin{figure}
	\includegraphics[width=0.85\linewidth]{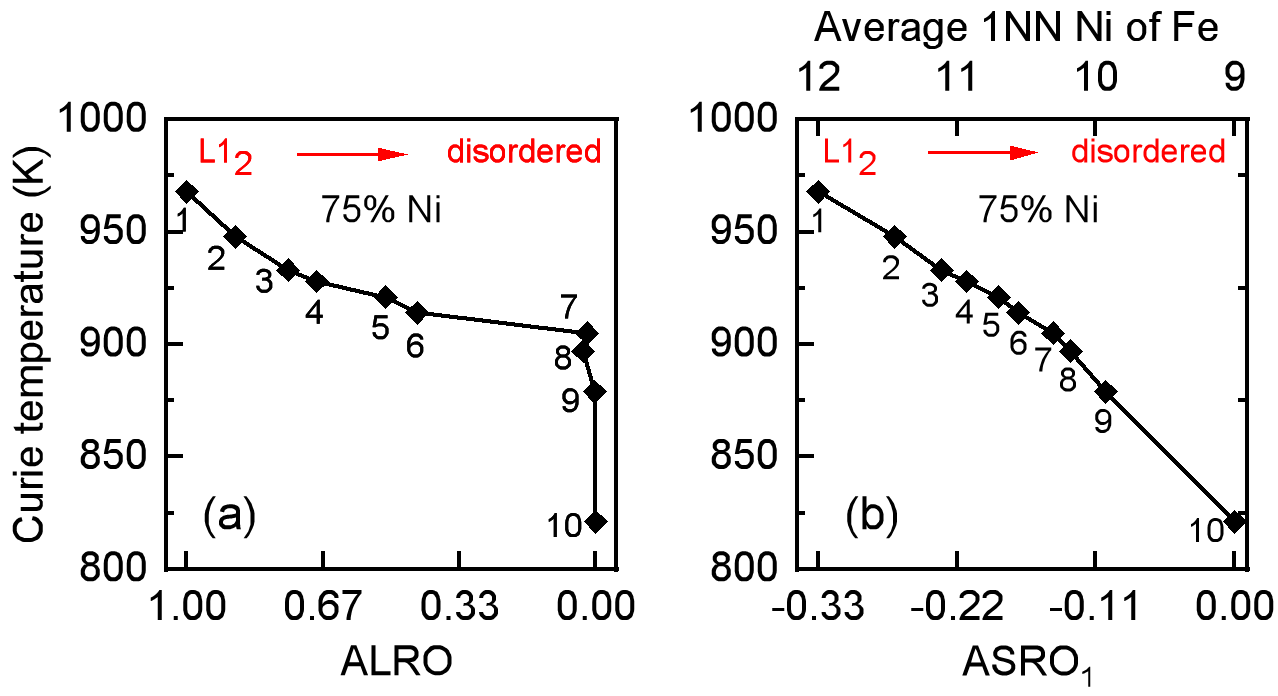}
	\includegraphics[width=0.85\linewidth]{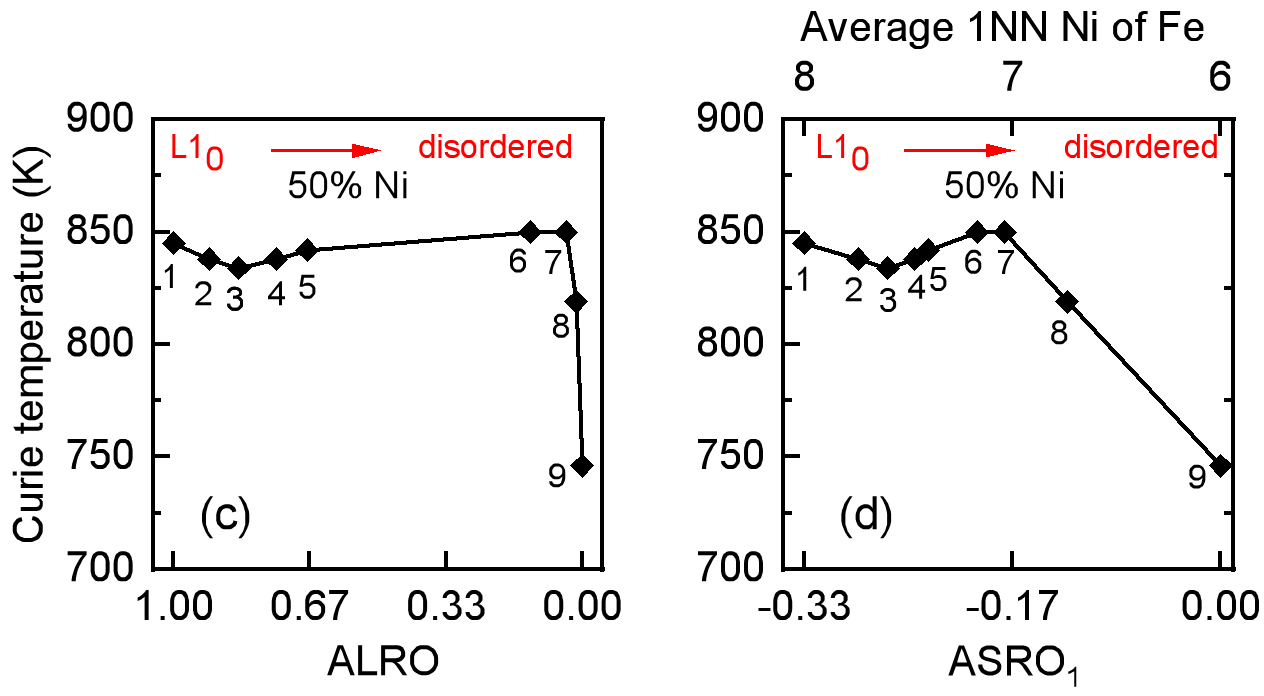}
	\caption{\label{fig:diff_LRO}(Color online) Curie temperatures as functions of ALRO and ASRO$_1$ in the structures with (a)(b) 75\% Ni and (c)(d) 50\% Ni. The numbers beside the symbols are used to label the same chemical configuration. The top X axis in (b) and (d) indicates the corresponding average number of first nearest Ni neighbours for an Fe atom. }
\end{figure}

Fig.~\ref{fig:diff_LRO} shows the Curie temperatures of these structures as functions of ALRO and the atomic short-range order of the first shell (ASRO$_1$). The ordering of the configurations are the same in the two figures (\textit{e.g.} the fifth data from the left in (a) and (b), or in (c) and (d), correspond to the same structure). The Curie temperature decreases from 968 K in the perfect L1$_2$ structure to 821 K in the completely disordered one. For structures with vanishing ALRO, their Curie temperatures can still differ by as much as about 80 K, due to the remaining ASRO. Indeed, it is found that the Curie temperatures have a rather linear dependence on ASRO$_1$ (Fig.~\ref{fig:diff_LRO}(b)). A similar investigation is also performed at 50\% Ni (Fig.~\ref{fig:diff_LRO}(c) and (d)). The Curie temperatures are found to be similar among the ordered structures (ALRO$>$0), whereas they are more sensitive to the variation in the ASRO$_1$ in the disordered state (ALRO=0).

\section{Vacancy formation properties in fcc Fe-Ni alloys}
\label{sec:vac}

\subsection{Accuracy of the model for vacancy-containing systems}
The present EIM is based on our previously developed EIMs of fcc Fe and Ni (namely the previous model parameters are kept), whose accuracy has been demonstrated in a previous study~\cite{Li2021a}. In the following, we validate the EIM description of vacancy-containing Fe-Ni alloys, by comparing its predicted vacancy formation energies with DFT results in the ordered and disordered structures.

Calculating vacancy formation energies in alloys from DFT is nontrivial, because chemical potentials in alloys cannot be obtained in a straightforward way as in pure systems. In an ordered phase with a dilute amount of point defects (e.g. vacancies and antisites), chemical potentials and point-defect formation energies can be calculated within the grandcanonical ensemble formalism~\cite{Mayer1995,Woodward2001} or the canonical ensemble formalism~\cite{Hagen1998,Mishin2000,DeKoning2002,Rogal2014}. Here we use the canonical ensemble formalism, which was first proposed by Hagen and Finnis~\cite{Hagen1998} and further developed by Mishin and Herzig~\cite{Mishin2000}, to compute vacancy and antisite formation energies. The results calculated using the DFT and EIM data are shown in Table~\ref{tab:E_f.ordered}.


First, it can be seen that the antisite formation energies are much lower than the vacancy formation energies in the Fe-Ni ordered structures. The Fe-Ni ordered structures are therefore the so-called antisite-disorder compounds, which have been studied in details by Mishin and Herzig~\cite{Mishin2000}. In particular, it is shown that the antisite formation energies in the two sublattices are equal in antisite-disorder compounds~\cite{Mishin2000}.

As shown in Table~\ref{tab:E_f.ordered}, there is a reasonable agreement between the DFT and EIM predictions of the vacancy formation energies in the Fe and Ni sublattices of L1$_2$-FeNi$_3$ and in the Ni sublattice of L1$_0$-FeNi, while the EIM result in the Fe sublattice of L1$_0$-FeNi is underestimated by 0.34 eV compared with the DFT one.


\begin{table}[htbp]
  \centering
  \caption{Vacancy and antisite formation energies (in eV) in the Fe and Ni sublattices calculated from DFT and EIM for the stoichiometric L1$_2$-FeNi$_3$ and L1$_0$-FeNi structures at the 0 K limit. V$_\text{Fe}$ and Ni$_\text{Fe}$ denote a vacancy and a Ni antisite in the Fe sublattice, respectively.}
  \label{tab:E_f.ordered}
\begin{ruledtabular}
        \begin{tabular}{ccccc}
          & \multicolumn{2}{c}{L1$_2$-FeNi$_3$} & \multicolumn{2}{c}{L1$_0$-FeNi} \\
          \cline{2-3} \cline{4-5}  
          & \multicolumn{1}{l}{DFT} & \multicolumn{1}{l}{EIM} & \multicolumn{1}{l}{DFT} & \multicolumn{1}{l}{EIM} \\
          \hline 
    V$_\text{Fe}$ & 1.392 & 1.300 & 1.897 & 1.561 \\
    V$_\text{Ni}$ & 1.593 & 1.666 & 1.847 & 1.794 \\
    Fe$_\text{Ni}$ & \multirow{2}[0]{*}{0.256}   & \multirow{2}[0]{*}{0.287}  & \multirow{2}[0]{*}{0.275}  &  \multirow{2}[0]{*}{0.253}\\
    Ni$_\text{Fe}$ & & & & 
    \end{tabular}
\end{ruledtabular}
\end{table}

In concentrated disordered alloys, it is customary to calculate the local vacancy formation energy at site $i$ as~\cite{Piochaud2014,Zhang2015a,Zhao2016,Zhang2021a}, $E_f^i$:
\begin{equation}
E_f^i = E_{\text{tot},\text{V}_\text{i}} -E_\text{tot,0} + \mu_{}
\label{eq:E_f^i}
\end{equation}
where $E_\text{tot,0}$ is the energy of the system without a vacancy, $E_{\text{tot},\text{V}_\text{i}}$ is the energy of the system with a vacancy at site $i$, and $\mu$ is the chemical potential of the removed atom in the system. In the DFT-SQS approach, $\mu$ is often calculated via the Widom substitution~\cite{Widom1963,Piochaud2014,Zhao2016}, which requires a large number of atom substitutions at different sites. Since our objective is to validate the present EIM, we may consider the arithmetic average of vacancy formation energy $<E_f^i>$, which can be readily obtained from DFT data without calculating $\mu$. Indeed, the relation $E_\text{tot,0}=N(x_A\mu_{A}+x_B\mu_{B})$ allows to eliminate $\mu_{A}$ and $\mu_{B}$ in $<E_f^i>$~\cite{Zhang2015a}:
\begin{equation}
\begin{split}
<E_f^i> = & x_A( <E_{\text{tot},\text{V}_\text{A}}> -E_\text{tot,0} + \mu_{A})+ \\
        & x_B( <E_{\text{tot},\text{V}_\text{B}}> -E_\text{tot,0} + \mu_{B}) \\
      = & x_A<E_{\text{tot},\text{V}_\text{A}}>+x_B<E_{\text{tot},\text{V}_\text{B}}> - \frac{N-1}{N} E_\text{tot,0}
\end{split}
\end{equation}
where $<E_{\text{tot},\text{V}_\text{A}}>$ is the average energy of the system with an $A$ atom removed. 

\begin{figure}
	\includegraphics[width=0.75\linewidth]{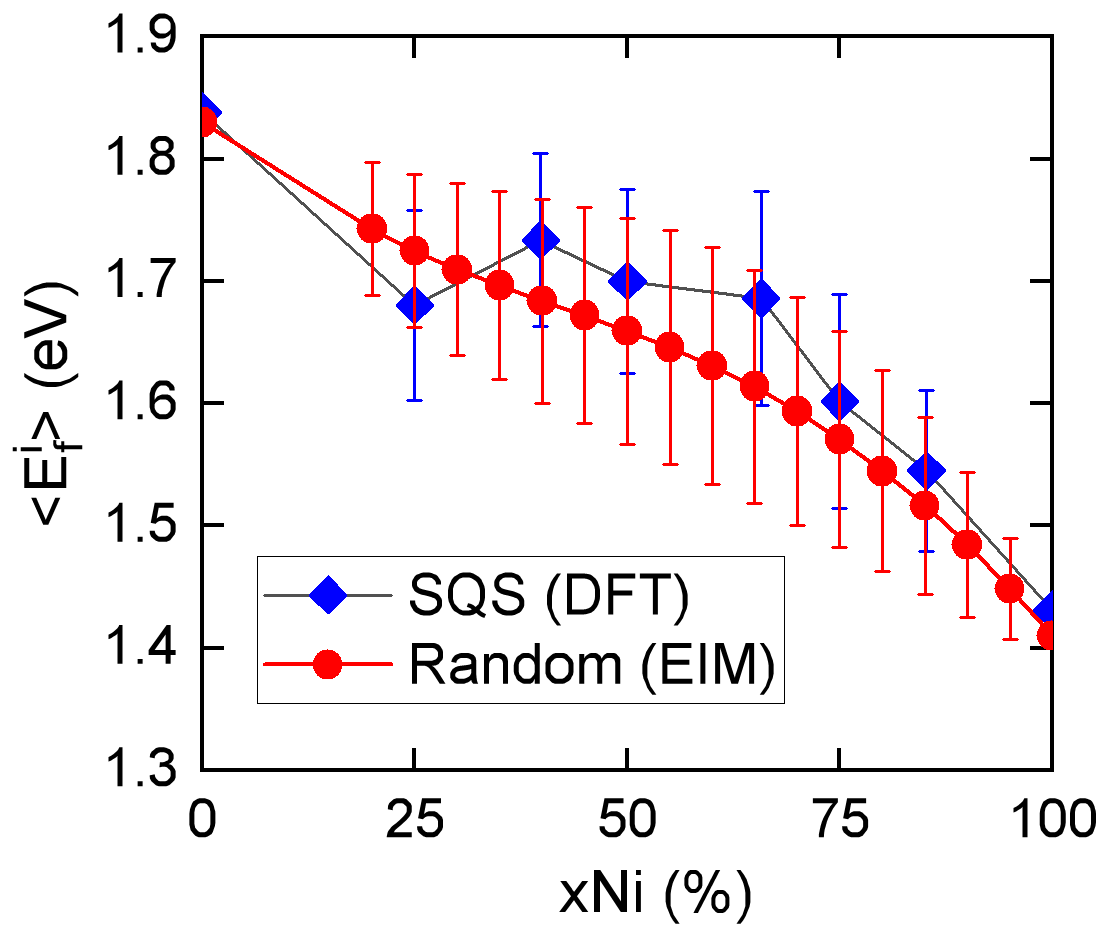}
	\caption{(Color online) Average vacancy formation energy as a function of Ni concentration for the random Fe-Ni structures in the magnetic ground state. The error bars denote the standard deviations of local vacancy formation energies.\label{fig:avg_Ef}}
\end{figure}

A comparison of $<E_f^i>$ in the random Fe-Ni structures between the DFT and EIM predictions is given in Fig.~\ref{fig:avg_Ef}. The DFT results are obtained with the 108-site SQSs. For each SQS, nine different Fe and Ni sites are considered to obtain $<E_{\text{tot},\text{V}_\text{Fe}}>$ and $<E_{\text{tot},\text{V}_\text{Ni}}>$, respectively. The EIM results are calculated in the 16384-site random structures in the magnetic ground state, and $<E_{\text{tot},\text{V}_\text{Fe}}>$ and $<E_{\text{tot},\text{V}_\text{Ni}}>$ are averaged over all the Fe and Ni sites, respectively. The DFT results suggest that $<E_{\text{tot},\text{V}_\text{Ni}}>$ decreases with increasing Ni concentration, which is also well reproduced by the EIM.

\subsection{Temperature dependence of vacancy formation properties}

\begin{figure}
	\includegraphics[width=\linewidth]{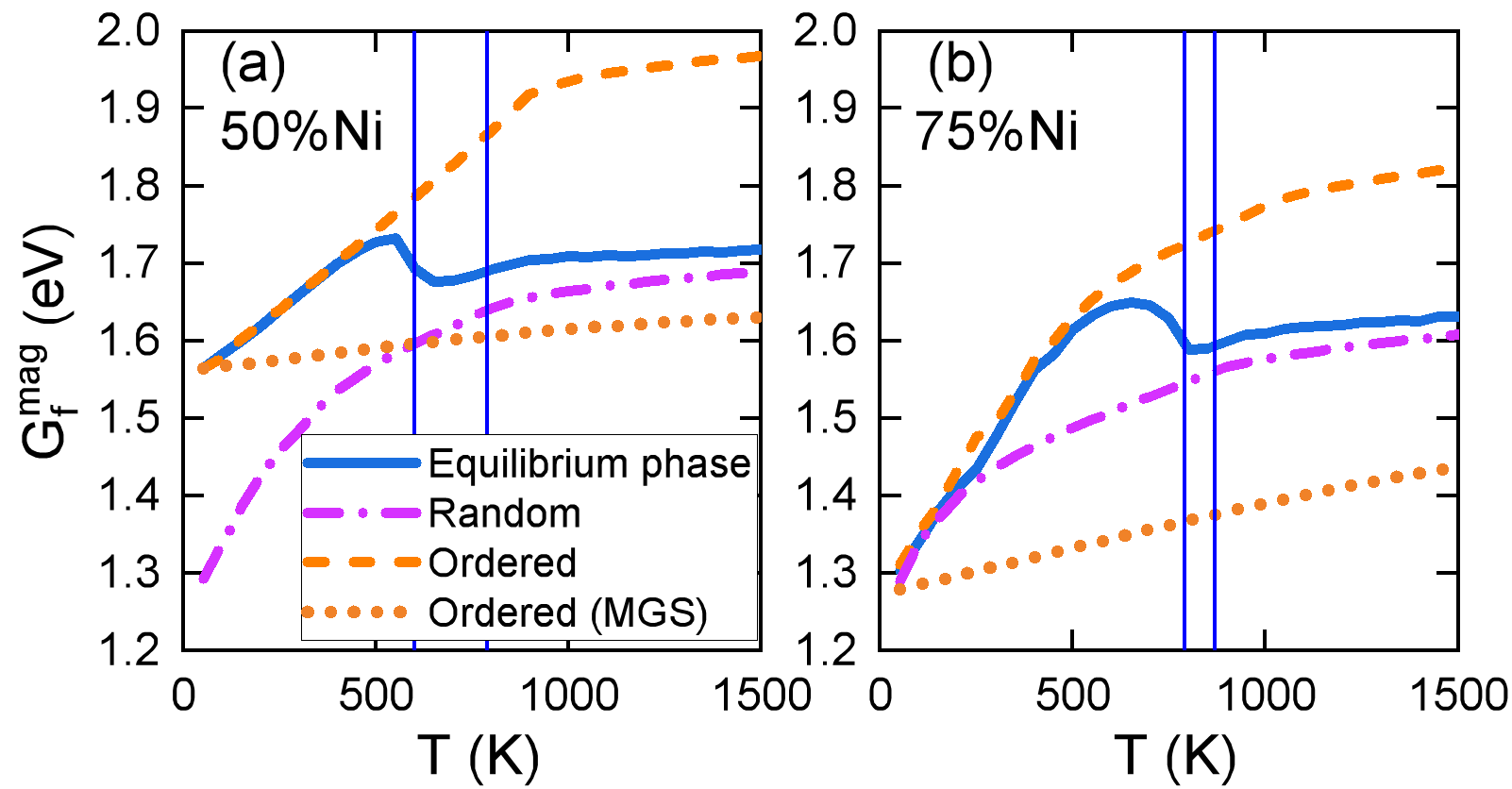}
	\caption{\label{fig:Gf_vs_T}(Color online) $G^\text{mag}_f$ as a function of temperature in the alloys with (a) 50 and (b) 75\% Ni. The solid lines are obtained in the equilibrium structures, and the vertical lines denote the corresponding chemical and magnetic transition temperatures. The dash-dotted (or dashed) lines are obtained in the random (or ordered) structures where the chemical order is frozen and only the magnetic order evolves with temperature. The dotted lines are obtained in the ordered structures in the magnetic ground state (both chemical and magnetic configurations are frozen).}
\end{figure}

The temperature evolution of $G^\text{mag}_f$ in the Fe-Ni alloys with 50\% and 75\% Ni, where the system undergoes successively the chemical and magnetic transitions with increasing temperature, is investigated and the results are shown in Fig.~\ref{fig:Gf_vs_T}. In the equilibrium phases, $G^\text{mag}_f$ first increase with increasing temperature, and then decrease abruptly across the chemical transition temperatures and finally increase slowly. 

This variation of $G^\text{mag}_f$ is clearly related to the changes of magnetic and chemical orders in the equilibrium phases. To separate these contributions, we calculate $G^\text{mag}_f$ in the structures where the chemical configurations are frozen while the magnetic configurations are equilibrated at each temperature. It can be seen that $G^\text{mag}_f$ in the equilibrium phases with 50\% and 75\% Ni follow closely those in the corresponding ordered structures up to 500 K and 600 K, respectively. This can be correlated with the previous results in Fig.~\ref{fig:order-disorder}, which show that these alloys remain fairly ordered up to 500 K and 600 K, with an ALRO$>$0.96. Near the chemical transition temperatures, $G^\text{mag}_f$ in the equilibrium phases deviate the trends in the ordered structures, but approach those in the disordered phases. As $G^\text{mag}_f$ are higher in the ordered structures than in the respective disordered ones, the chemical transitions thus lead to a decrease in $G^\text{mag}_f$. 

Fig.~\ref{fig:Gf_vs_T} indicates that $G^\text{mag}_f$ in the ordered and disordered structures increase with increasing temperature. Such variations are related not only to magnetic excitations, but also to the changes of weights in the local vacancy formation energies. Indeed, even if the structures are chemically and magnetically frozen, $G^\text{mag}_f$ still tend to increase with temperature. For example, $G^\text{mag}_f$ in L1$_2$-FeNi$_3$ with the magnetic ground state can be calculated as
\begin{align}
\label{eq:Gf2}
G^\text{mag}_f= & -k_\text{B}T \, \ln [ 0.25 \cdot \exp (-\dfrac{E_f^\text{Fe-lat}}{k_\text{B}T}) \notag \\
& + 0.75 \cdot \exp (-\dfrac{E_f^\text{Ni-lat}}{k_\text{B}T}) ]
\end{align}
where $E_f^\text{Fe-lat}$ and $E_f^\text{Ni-lat}$ are the vacancy formation energies in the Fe and Ni sublattices given in Table~\ref{tab:E_f.ordered}, respectively. As suggested by the expression, the lower $E_f^\text{Fe-lat}$ has a dominant weight in the evaluation of $G^\text{mag}_f$ at low temperatures, but $E_f^\text{Fe-lat}$ and $E_f^\text{Ni-lat}$ eventually have similar weights at high temperatures. As a result, $G^\text{mag}_f$ in L1$_2$-FeNi$_3$ increases from $E_f^\text{Fe-lat}$ at low temperature, to the arithmetic average of all local vacancy formation energies at the high-temperature limit, namely:
\begin{align}
\label{eq:Gf_inf}
G^\text{mag}_f = 0.25 E_f^\text{Fe-lat} + 0.75 E_f^\text{Ni-lat}, \; \text{for} \; T \! \rightarrow{} \!\!+\! \infty{}
\end{align}
The dotted lines in Fig.~\ref{fig:Gf_vs_T} denote $G^\text{mag}_f$ in L1$_0$-FeNi and L1$_2$-FeNi$_3$ in the respective magnetic ground states. Comparing these results to $G^\text{mag}_f$ in the same ordered structures but with the equilibrium magnetic configurations, it can be concluded that the increase of $G^\text{mag}_f$ in the latter cases is mainly due to the magnetic excitations. More specifically, it is primarily related to the transversal spin fluctuations, since the longitudinal spin fluctuations in the ordered structures are found to be relatively weak below the Curie temperatures. With additional DFT calculations, we confirm the increasing behaviour of $G^\text{mag}_f$ predicted by the EIM, though the magnetic disordering effects are found to be somehow exaggerated by the EIM~\cite{suppl_mat}.

\subsection{Concentration dependence of vacancy formation properties}

The predicted magnetic free energies of the Ni-vacancy binding in Fe and the Fe-vacancy binding in Ni are shown in Table~\ref{tab:Gb}. The solute-vacancy interactions in Fe and Ni in the magnetic ground state are quite weak, being marginally attractive and repulsive, respectively. The magnetic transition in fcc Fe and Ni changes the binding magnetic free energy only slightly, by less than 0.04 eV. According to these results, the Ni-V interaction in fcc Fe and the Fe-V interaction in Ni are not significant at any temperature.

\begin{table}[htbp]
  \centering
  \caption{Solute-vacancy binding free energy (in eV) in fcc Fe and Ni in the magnetic ground state (MGS) and the PM state. The binding energies in the intermediate temperature range lie between the values of the MGS and PM states. In our convention, a positive value indicates an attraction between the vacancy and the solute.}
\begin{ruledtabular}
    \begin{tabular}{ccccc}
          & \multicolumn{2}{c}{Ni+V in fcc Fe} & \multicolumn{2}{c}{Fe+V in fcc Ni} \bigstrut\\
\cline{2-5}          & 1NN   & 2NN   & 1NN   & 2NN \bigstrut\\
    \hline
    MGS    & 0.02  & 0.03  & -0.05 & -0.03 \bigstrut[t]\\
    PM (1500 K) & 0.01  & -0.03 & -0.04 & -0.07 \bigstrut[b]\\
    \end{tabular}
\end{ruledtabular}
  \label{tab:Gb}
\end{table}

\begin{figure}
	\includegraphics[width=0.8\linewidth]{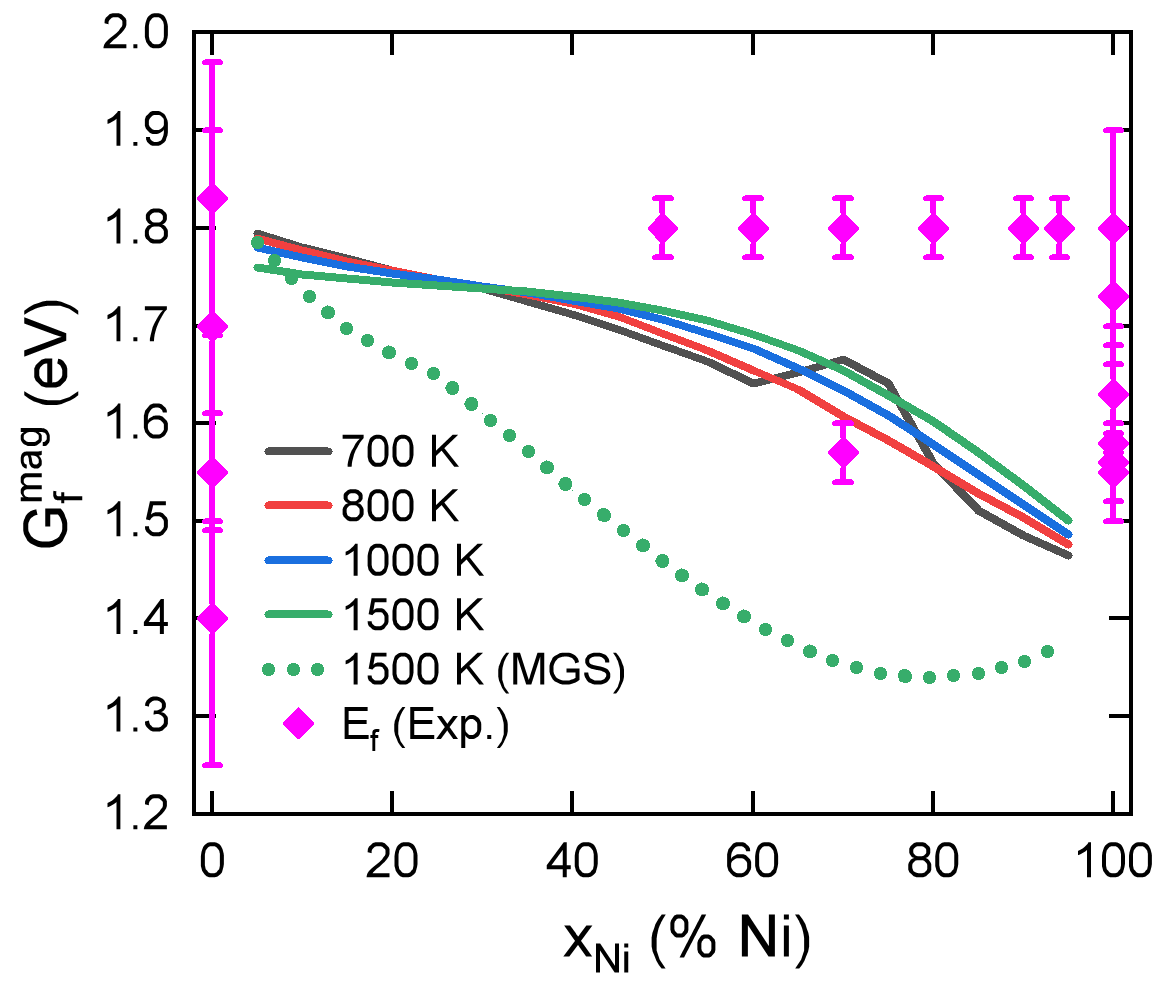}
	\caption{\label{fig:Gf_vs_xNi}(Color online) The predicted $G^\text{mag}_f$ as a function of Ni concentration at several temperatures, compared to the experimental vacancy formation energies (fcc Fe~\cite{Schaefer1977,Schulte1978,Kim1978,Matter1979}, fcc Ni~\cite{Wycisk1978,Smedskjaer1981,Wolff1997,Scholz2001,Troev1989}, fcc Fe-Ni alloys~\cite{Chambron1974,Caplain1977}). The solid lines denote the results obtained in the equilibrium spin-atom structures, whereas the dotted line denotes the results obtained in the chemically disordered structures in the magnetic ground states (MGS), which are collinear FM above 25\% Ni and non-collinear below 25\% Ni~\cite{suppl_mat}.}
\end{figure}

The predicted concentration dependence of $G^\text{mag}_f$ in fcc Fe-Ni alloys at several temperatures is shown in Fig.~\ref{fig:Gf_vs_xNi}. According to the calculated phase diagram, the equilibrium phases above 770 K are solid solutions for all compositions. As shown in Fig.~\ref{fig:Gf_vs_xNi}, the computed $G^\text{mag}_f$ at 800 K and above tends to decrease with increasing Ni concentration. This trend is also observed in the curve of $G^\text{mag}_f$ at 700 K, except in the composition range of 60-80\% Ni where the alloys have an L1$_2$ ordered structure. 

As shown in Fig.~\ref{fig:Gf_vs_xNi}, $G^\text{mag}_f$ increase weakly with increasing temperature in the disordered structures with more than 30\% Ni, while the trend is reversed in the disordered structures below 30\% Ni. This can be correlated with our previous results which suggest that $G^\text{mag}_f$ in fcc Fe and in Ni decreases and increases with increasing temperature, respectively~\cite{Li2021a}.


\begin{figure}
	\includegraphics[width=0.75\linewidth]{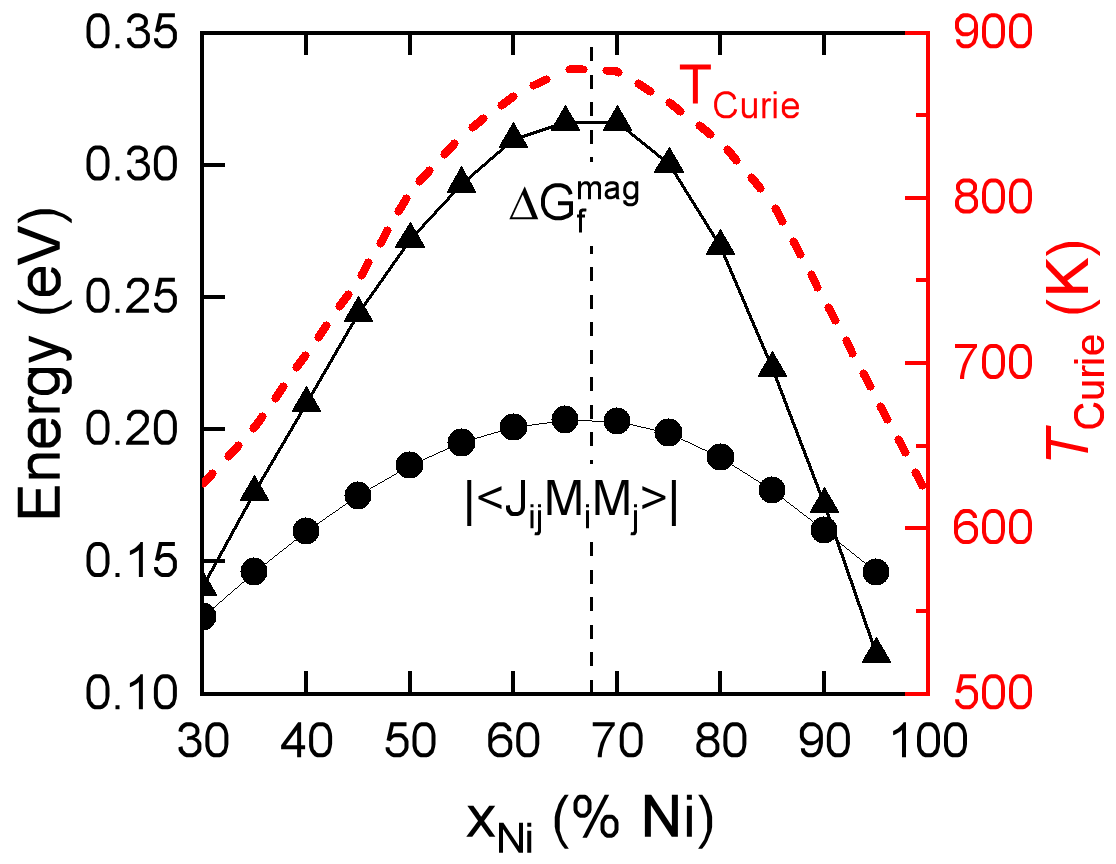}
	\caption{\label{fig:GS_spin}(Color online) Difference between $G^\text{mag}_f$ in the disordered structures at 1500 K with the equilibrium magnetic states and with the magnetic ground states, the average magnetic exchange interaction energies $|<J_{ij}\vect{M}_i \vect{M}_j>|$ and the Curie temperatures in the disordered structures.}
\end{figure}

The small variation of $G^\text{mag}_f$ with temperature in Fig.~\ref{fig:Gf_vs_xNi} does not mean that magnetism has no impact on $G^\text{mag}_f$. Indeed, there is already a large extent of magnetic disorder at temperatures where the equilibrium chemical configurations are disordered. Therefore, the effects of thermal spin fluctuations are less significant in the disordered alloys with the equilibrium magnetic states. In Fig.~\ref{fig:Gf_vs_xNi}, we also show $G^\text{mag}_f$ calculated in the disordered structures in the respective magnetic ground states. They are much lower than the alloys with the equilibrium magnetic states in the concentrated composition range. As presented in Fig.~\ref{fig:GS_spin}, the difference between the two curves of $G^\text{mag}_f$ at 1500 K reaches a maximum of 0.32 eV around 65\% Ni, where the Curie temperature is also the highest. The latter is a sign of the magnitude of the magnetic interaction energy, which is also the strongest around 65\% Ni according to our model. Indeed, it is shown that the difference between $G^\text{mag}_f$ in the paramagnetic state and the ground state is closely related to the magnetic interaction energy~\cite{Li2021a}. 


Finally, the calculated $G^\text{mag}_f$ are compared to the experimental vacancy formation energies $E_f$ in Fig.~\ref{fig:Gf_vs_xNi}. We note that the calculated $G^\text{mag}_f$ and $E_f$ are similar above 1000 K, which is in the range of temperatures where the measurements of $E_f$ were performed. To the best of our knowledge, the measurements of $E_f$ in fcc Fe-Ni alloys have been reported only by Caplain and Chambron using magnetic anisotropy measurements~\cite{Chambron1974,Caplain1977}. In their first study, measurements were performed in the disordered Fe-Ni samples with 70\% Ni quenched from between 873 and 973 K, and $E_f$ was found to be 1.57 eV~\cite{Chambron1974}. In their subsequent study in the disordered samples with 50\% to 94\% Ni quenched from above the chemical transition temperature, $E_f$ were found to be 1.80 eV regardless of the composition~\cite{Caplain1977}. It is difficult to draw a definitive conclusion regarding the concentration dependence or the values of $E_f$ based solely on these two experiments, which could have large experimental uncertainty as in the cases of pure fcc Fe and Ni. On the other hand, our results of $G^\text{mag}_f$ fall between $E_f$ from these two sets of measurements, and they are within the uncertainty of the available experimental data over the whole concentration range. It would be useful to have further experimental investigations to clarify the validity of the current predictions.

\section{Conclusion}
\label{sec:conclu}

Phase stability and vacancy formation in fcc Fe-Ni alloys are investigated for a broad composition-temperature range, using an effective interaction model (EIM) combined with on-lattice Monte Carlo simulations.

Parametrized on DFT data only, the present EIM enables a good prediction of the experimental magnetic and chemical transition temperatures in the fcc Fe-Ni alloys over the whole range of composition. Compared with magnetic excitations, lattice vibrations show a larger impact on the chemical order-disorder transitions. The predicted fcc phase diagram is compared to the most recent CALPHAD assessment, showing an overall good agreement. In particular, the EIM predicts a phase separation in the disordered alloys around 10-40\% Ni and 570-700 K, which is shown to be magnetically driven. In addition, the magnetic state has a strong influence on the chemical order-disorder transition temperature, which can differ by up to 170 K. The Curie temperature is sensitive to both atomic long-range and short-range orders, and tends to increase with increasing chemical ordering.

Vacancy formation magnetic free energy $G^\text{mag}_f$ in fcc Fe-Ni alloys is studied as a function of temperature and composition. It is worth noting that the temperature evolution of $G^\text{mag}_f$ in the magnetic alloys cannot be described by the Ruch model~\cite{Ruch1976} or the Girifalco model~\cite{Girifalco1964} due to the simultaneous evolution of magnetic and chemical degrees of freedom. We find that magnetic disorder leads to an increase of $G^\text{mag}_f$ while chemical disorder has the opposite effect. In the solid solutions, $G^\text{mag}_f$ tends to decrease with increasing Ni concentration. Our results reveal that the effects of magnetic excitations and transitions on vacancy formation properties are much more significant in concentrated Fe-Ni alloys than in pure Fe and Ni, due to the strong magnetic interaction in the concentrated alloys as revealed in the concentration dependence of Curie temperatures. 


\begin{acknowledgments}
This work was performed using DARI-GENCI resources under the A0090906020 and A0110906020 projects, and the CINECA-MARCONI supercomputer within the SISTEEL project. K.L. is supported by the CEA NUMERICS program, which has received funding from the European Union's Horizon 2020 research and innovation program under the Marie Sk\l{}odowska-Curie grant agreement No. 800945.
\end{acknowledgments}

\bibliography{library} 

\end{document}